\documentclass[universe,article,accept,oneauthor,dvi2pdf,10pt,a4paper]{mdpi}

\firstpage{1}
\articlenumber{x}
\doinum{10.3390/------}
\pubvolume{3}
\pubyear{2017}
\copyrightyear{2016}
\externaleditor{Academic Editors: Valerio Bozza, Francesco De Paolis and Achille A. Nucita
} 
\history{Received: 24 November 2016; Accepted: 22 December 2016; Published: date} 
\usepackage{soul}
\usepackage{upgreek}
\usepackage{enumitem}
\usepackage{subfigure}
\usepackage{booktabs}
\usepackage{multirow}
\usepackage{longtable}
\usepackage{microtype}
\newcommand\myurl[1]{\changeurlcolor{black}\url{#1}\changeurlcolor{blue}}

\usepackage{graphicx}
\usepackage{ogonek}
\usepackage{amstext}
\usepackage{url}
\usepackage{color}

\Title{Gravitational Lensing of Rays through the Levitating Atmospheres of Compact Objects}

\Author{Adam Rogers} 
\AuthorNames{Adam Rogers}

\address[1]{Department of Physics and Astronomy, University of Manitoba, Winnipeg, MB R3T 2N2, Canada; rogers@physics.umanitoba.ca}

\corres{Correspondence: rogers@physics.umanitoba.ca}

\abstract{Electromagnetic rays travel on curved paths under the influence of gravity. When~a~dispersive optical medium is included, these trajectories are frequency-dependent. In~this work we consider the behaviour of rays when a~spherically symmetric, luminous compact object described by the Schwarzschild metric is surrounded by an~optically thin shell of plasma supported by radiation pressure. Such levitating atmospheres occupy a~position of stable radial equilibrium, where~radiative flux and gravitational effects are balanced. Using general relativity and an~inhomogeneous plasma we find the existence of a~stable circular orbit within the atmospheric shell for low-frequency rays. We explore families of bound orbits that exist between the shell and the compact object, and identify sets of novel periodic orbits. Finally, we examine conditions necessary for the trapping and escape of low-frequency radiation.}


\begin{document}

\section{Introduction}
\label{sec:intro}

  Novel behaviours for electromagnetic rays result from the combination of gravitational lensing and the optical effects from a~dispersive plasma medium \cite{perlickGR, review}. The joint effects of gravitation, refraction~and dispersion have been studied using homogenous \cite{BKT09, BKT10} and inhomogenous plasma distributions \cite{TBK13} on a~variety of scales \cite{mao14, shadow1} and geometries \cite{plasmaLensingKerr, shadow2, shadow3}. \textcolor{black}{Bound orbits of rays under the effect of gravity and plasma have also been discussed \cite{kulsrudLoeb, TBK13}.} The effect of a~power-law distribution of transparent plasma on the pulse profiles of a~compact object (CO) have been shown to produce frequency-dependent shifts \cite{rogers15}, and trapping of rays \cite{rogersOrbits}.

  In the environment of a~highly luminous neutron star (NS), a~distribution of plasma is strongly affected by both radiation pressure and gravity \cite{abram, lambMiller, ECSDetail1}. In the Newtonian picture, the radiative force decreases as $r^{-2}$, and therefore if radiation overpowers gravity at a~particular radius it is true for all space. However, the situation described by general relativity is more subtle. In Schwarzschild space-time the radiative force changes at a~faster rate than gravitational effects \cite{ECSstable2, stahl13}. Thus, there~is only a~single radius at which the effects of gravity and radiation on a~test particle are balanced. This~position of radial equilibrium defines a~surface called the Eddington capture sphere (ECS \cite{ECSstable1, ECSstable2}). At this radius, a~stable shell of plasma can collect, detached from the NS surface \cite{atmShellOsc}. These levitating atmospheric shells have been studied in optically thin and optically thick conditions \cite{atmThin, atmThick}.

  In this work, we assume that the optically thin levitating atmosphere around a~CO can be treated as a~transparent plasma shell and that radiation is free to propagate between the stellar surface and the shell. We calculate the resulting frequency-dependent ray trajectories under the effects of gravitation from the CO as well as refraction and dispersion from the plasma shell. In Section \ref{sec:theory} we discuss the theoretical aspects of levitating atmospheres and the ray-tracing procedure. In Section \ref{sec:numerical} we present our numerical results. \textcolor{black}{Section \ref{sec:discussion} provides discussion of our calculations and the general results that we have obtained, as well as example density functions}. Finally, we summarize our findings in Section \ref{sec:conclusions}.

\section{Theory}
\label{sec:theory}

Let us use the set of units in which $G=c=\hbar=1$. We assume spherical symmetry, and without loss of generality work in the equatorial plane $\theta=\pi/2$. The Schwarzschild line element is
\begin{equation}
\text{d}s^2=-A(r)\text{d}t^2+\frac{\text{d}r^2}{A(r)}+r^2 \text{d}\phi^2
\end{equation}
with
\begin{equation}
A(r) = 1-\frac{2M}{r}.
\end{equation}

As a~first approach, let us treat the plasma shell as a~non-magnetized, inhomogenous optically-thin medium with index of refraction of the form
\begin{equation}
n^2(r) = 1 - \frac{\omega_p(r)^2}{\omega^2}
\label{nRefr}
\end{equation}
\textcolor{black}{with ray frequency $\omega$} and local plasma frequency
\begin{equation}
\omega_p^2 = \frac{4 \pi q^2 N_{0}}{m} N(r) = k N(r)
\label{plasmaFreq}
\end{equation}
where the plasma particles have charge $q$ and mass $m$, and $N(r)$ describes the number density distribution of the plasma, scaled by the maximum $N_0$. On the right hand side we have consolidated the coefficients into the constant $k$. For the remainder of our numerical work we use a~simple exponential function to describe the shell density
\begin{equation}
N(r)=\exp \left[ -\frac{ (r - r_{0})^2 }{ \sigma^2 } \right].
\label{NDef}
\end{equation}
with the maximum \textcolor{black}{located at a~distance $r_\text{0}>R$, where $R$ is the radius of the CO.} Despite these assumptions, our conclusions will be general, and will not depend on the exact details of the density function. The numerical examples in the body of the text use Equation \eqref{NDef}, however~we consider other density functions in Section \ref{sec:discussion}. In fact, any shell-like density profile will generate qualitatively similar results. Throughout the text we will refer to the plasma density as $N$, and suppress the radial dependence for simplicity.

\textcolor{black}{The Hamiltonian for an~electromagnetic ray under the effect of gravity and an~optical medium was first considered by Synge \cite{synge}. This result was subsequently specialized to non-homogenous plasma \cite{BKT09, perlickGR}, which we state using Equations \eqref{nRefr}--\eqref{NDef},}
\begin{equation}
H=\frac{1}{2}\left( g^{ij}p_i p_j + k N \right).
\label{eq:hamiltonian}
\end{equation}
The equations of motion are
\begin{equation}
\frac{\text{d}x^i}{\text{d}\lambda}=\frac{\partial H}{\partial p_i} = g^{ij}p_j
\end{equation}
and
\begin{equation}
\frac{\text{d}p_i}{\text{d}\lambda}=-\frac{\partial H}{\partial x^i}= -\frac{1}{2}g^{jk}_{,i}p_j p_k -\frac{k}{2}\left( N \right)_{,i}.
\end{equation}

\textcolor{black}{The derivatives of the $t$ and $\phi$ momenta vanish, providing the constants of motion}
\begin{equation}
p_t = - E =- \omega_\infty
\label{pT}
\end{equation}
\textcolor{black}{where $\omega_\infty$ is the ray frequency an~observer at infinity would detect provided the ray escapes,} and
\vspace{12pt}
\begin{equation}
p_\phi = L
\label{pPhi}
\end{equation}
\textcolor{black}{where $L$ is the angular momentum. Since we restrict the trajectories to the equatorial plane,} the~momentum $p_\theta$ vanishes. With this restriction the indices $i$, $j$ and $k$ each take the values $t$, $r$ and $\phi$. For a~static medium the effective redshift relationship is given by
\begin{equation}
\omega(r)=\frac{\omega_\infty}{A(r)^\frac{1}{2}}.
\label{redshiftDef}
\end{equation}
The corresponding coordinate derivatives are
\begin{equation}
\frac{\text{d} t}{\text{d}\lambda} = \frac{E}{A(r)}
\end{equation}
\begin{equation}
\frac{\text{d} \phi}{\text{d}\lambda} = \frac{L}{r^2}.
\label{dphidlambda}
\end{equation}

For spherical symmetry, \textcolor{black}{there is no orbital motion out of the $\theta=\pi/2$ plane.} The radial momentum is given by Equation \eqref{eq:hamiltonian} vanishing,
\begin{equation}
p_r= \pm \frac{L}{A(r)} \left[ \frac{E^2}{L^2} - A(r)\left( \frac{1}{r^2} + \frac{k N}{L^2} \right) \right]^{ \frac{1}{2} }
\label{pR}
\end{equation}
with the sign of $p_r$ denoting a~radially infalling ray as $p_r<0$ and an~outgoing ray with $p_r>0$. The~derivative of the radial momentum is
\begin{equation}
\frac{\text{d} p_r}{\text{d}\lambda}= -\frac{M}{r^2 A^2(r)} E^2 - \frac{M}{r^2} p_r^2 + \frac{L^2}{r^3} - \frac{k}{2} \frac{\text{d} N }{\text{dr}} ,
\end{equation}
giving the radial coordinate derivative
\begin{equation}
\frac{\text{d} r}{\text{d}\lambda} = A(r) p_r.
\label{dr}
\end{equation}
Finally, we find the change in angle as a~function of radius using Equations \eqref{dphidlambda}, \eqref{pR} and \eqref{dr},
\begin{equation}
\frac{\text{d}\phi}{\text{dr}} = \pm \frac{1}{r^2\left( \frac{E^2}{L^2}n(r)^2-\frac{A(r)}{r^2} \right)^\frac{1}{2}}.
\label{dphidr}
\end{equation}

The effective potential that includes the effect of the plasma is
\begin{equation}
V(r)=\left( 1-\frac{2M}{r}\right)\left(\frac{L^2}{r^2} + kN \right),
\end{equation}
with first and second derivatives
\begin{equation}
\begin{array}{ll}
\frac{\text{d}V}{\text{dr}}= & -\frac{2L^2}{r^3} \left( 1-\frac{3M}{r} \right)  \\
 & + k \left[ \left(1-\frac{2M}{r} \right)\frac{\text{d}N}{\text{dr}} + \frac{2M}{r^2}N \right]
\end{array}
\label{diffV1gen}
\end{equation}
\begin{equation}
\begin{array}{ll}
\frac{\text{d}^2 V(r)}{\text{dr}^2} = & \frac{6L^2}{r^4}\left(1-\frac{4M}{r} \right) \\
 & + k \left[ \left( 1- \frac{2M}{r} \right) \frac{\text{d}^2 N}{\text{dr}^2} + \frac{4M}{r^2} \frac{\text{d}N}{\text{dr}} - \frac{4M}{r^3}N \right].
\end{array}
\label{diffV2gen}
\end{equation}
These general expressions with power-law density reproduce the corresponding equations from~\cite{rogers15}. Circular orbits are found at radial distances where the derivative of the effective potential vanishes \cite{v1, ve1}. \textcolor{black}{We call these extremal radii $r_\text{e}$, where $e$ takes the labels $s$ for a~stable orbit and $u$ for an~unstable orbit. Stable minima further~require}
\vspace{12pt}
\begin{equation}
\left. \frac{\text{d}^2V(r)}{\text{dr}^2} \right|_{r_\text{s}} > 0.
\end{equation}
Solving for the corresponding $L$ with $\text{d}V/\text{dr}=0$ gives the angular momentum required to produce a~circular orbit
\begin{equation}
\textcolor{black}{L_\text{e}^2= \frac{r_\text{e}^4 k^\frac{1}{2}}{2\left( r_\text{e}-3M \right)}\left[ \left(1-\frac{2M}{r_\text{e}}\right)\left.\frac{\text{d}N}{\text{dr}}\right|_{r_\text{e}} + \frac{2M}{r_\text{e}^2}N_\text{e} \right]^\frac{1}{2}}
\end{equation}
with $N_\text{e}=N(r_\text{e})$.

  Levitating spherical shell-type atmospheres around COs are physically motivated and have an~established theoretical foundation in the literature \cite{abram, atmThick}. To explore the basics of these solutions, \textcolor{black}{let us follow the approach in \cite{atmThin} and write the observed stellar luminosity as $\ell_\infty$}. Then the local stellar luminosity as a~function of radial distance $r$ from the CO is given by
\begin{equation}
\textcolor{black}{\ell(r)=\frac{\ell_\infty}{A(r)}.}
\label{stellR}
\end{equation}
The observed Eddington luminosity is
\begin{equation}
\textcolor{black}{\ell_\text{Edd}=\frac{4 \pi M m}{\sigma_\text{T}} }
\end{equation}
where $m$ is the plasma particle mass and $\sigma_\text{T}$ is the cross-section for Thomson scattering. However, the Eddington luminosity can also be used to define a~local critical luminosity required to produce a~radiative force which balances gravity at a~distance $r$ \cite{ECSDetail1}. The local critical luminosity is
\begin{equation}
\textcolor{black}{\ell_\text{c}(r)=\frac{\ell_\text{Edd}}{A^\frac{1}{2}(r)}.}
\label{eddR}
\end{equation}
The radius at which the stellar luminosity $\ell$ is equal to the critical luminosity $\ell_c$ defines a~spherical surface around the CO called the Eddington capture sphere \cite{ECSDetail1, atmThin} with radius
\begin{equation}
r_\text{ECS}=\frac{2M}{1-\lambda^2},
\end{equation}
where the ratio of the observed luminosity and the Eddington luminosity is
\begin{equation}
\textcolor{black}{\lambda=\frac{\ell_\infty}{\ell_\text{Edd}}. }
\end{equation}
Plasma particles near the CO are subject to radiation drag and have their angular momentum reduced \cite{ECSstable3}. These particles gather on the ECS since it is an~equilibrium position in the radial direction~\cite{ECSstable1, ECSstable2} and is stable against radial oscillations \cite{atmShellOsc}.

  The stability of the ECS allows for the formation of a~levitating shell of plasma. For larger values of the luminosity ratio, $0<\lambda<1$, these levitating atmospheres are separated from the NS surface with no significant density between. Analytical density profiles have been constructed for isothermal and polytropic levitating atmospheres in the optically thin case \cite{atmThin}. The polytropic atmospheres have thickness that depends on the temperature $T$, with higher $T$ producing more distended atmospheres. The optically thick case has also been investigated numerically \cite{atmThick}. The radiative equilibrium condition has been used in the analysis of photospheric expansion X-ray bursts from NSs \cite{atm1, atm2, Xray1, Xray2}, and the effect of a~levitating atmosphere may also contribute to the nature of these bursts \cite{ECSstable3, atmThin}.

  Our assumed density $N$ (Equation \eqref{NDef}) approximates the rapid drop-off from the peak density $r_{0}=r_\text{ECS}$ of the levitating polytropic fluid shells \cite{atmThin} while remaining analytically simple. However, our conclusions are qualitatively insensitive to the particulars of $N$, provided a~maximum external to the stellar surface. Thus, we are not concerned with the exact details that lead to the levitating atmosphere solution and rather seek to capture the general character of these plasma shells in our numerical examples.

  We will examine the trajectories of electromagnetic rays by ray-tracing. We assume a~ray of energy $E$ and angular momentum $L$ begins at an~initial position $(t, r, \phi)$, directed inward or outward depending on the choice of sign for $p_r$. We then solve Equations \eqref{pT}--\eqref{dr} numerically to find the path of the ray under the influence of gravity and the effect of the plasma given by the density $N$ in Equation \eqref{NDef}. The integration of the equations of motion is ceased when the trajectory intersects the stellar surface or escapes to a~significant distance ($r \geq 100 R$). For bound rays, we follow the orbits for an~arbitrary number of time-steps, chosen to give a~few periastron-apastron pairs.

\section{Numerical Results: A~Variety of Novel Orbits}
\label{sec:numerical}

To initialize our numerical ray-tracing routine, we select the physical parameters that describe our CO and the levitating atmospheric shell. We choose the mass of the CO as $M=1.4 M_\odot$. To maximize the gravitational effects, we use an~extremely relativistic compactness ratio of $R/M=1.6$, at the stiff end of the nuclear EoS \cite{rogersOrbits, NSEoS}. We choose a~relatively high ratio of stellar to Eddington luminosity of $\lambda=0.9$, which gives the center of the shell at $r_\text{ECS}=10.5 M$. Finally, we estimate the thickness of the shell using the approximate formula from \cite{atmThin},
\begin{equation}
H \approx 2 \times 10^{-3} \frac{2M}{ \left( 1-\lambda \right)^{2} }
\end{equation}
which gives $H \approx 1$ km for our choices of parameters. We then set $\sigma=H$ and $r_0=r_\text{ECS}$ in our test function, given in Equation \eqref{NDef}.

  The equations of motion show us that the constant $k$ sets the frequency at which plasma effects become relevant to the ray trajectories, but the dynamics of the rays are unchanged provided we redefine the momenta $p'_i=p_i/\sqrt{k}$, as well as the affine parameter $\lambda'=\sqrt{k} \lambda$. Thus, in our numerical examples we simply set $k=1$ and discuss the relevant frequency scale in Section \ref{sec:discussion}.

\subsection{A Stable Circular Orbit for Electromagnetic Rays}
\label{sub:circular}

  When an~optically thin levitating atmosphere with a~refractive index is considered, the effective potential is significantly more complicated than the vacuum case and allows for a~variety of circular orbits. As a~numerical example, let us consider a~ray that has a~frequency $\omega_\infty=1$. Such a~ray with angular momentum $L_\text{s}=15.1689$ has a~stable circular orbit at $r_\text{s}=13.6468$. This stable circular orbit is well within the radius of the ECS, $r_\text{ECS}=14.7368$. The unstable circular orbit radius for electromagnetic rays at the potential maximum, $r_\text{u}=14.7194$, is shifted inward slightly from the ECS radius due to the contribution from the angular momentum term in the effective potential.

  Generally, any shell-like plasma atmosphere with a~density profile $N$ that has a~maximum at $r_0>R$ and which is physically realistic (decreases as a~function of $r$ in both directions $r>r_{ECS}$ and $r<r_{ECS}$ and which vanishes at infinity) will produce a~minimum in the effective potential, resulting~in a~stable circular orbit. The existence of this stable circular orbit does not require any additional symmetry arguments be imposed on the shell density, provided the previous conditions are satisfied. For example, an~asymmetrical density profile was found for an~isothermal levitating atmosphere \cite{atmThin}, though this is not a~physically realistic solution since the density does not vanish at infinity. However, such a~density profile would still produce a~stable circular orbit within $r_\text{ECS}$.

\subsection{Periodic Orbits}
\label{subsec:periodic}

\begin{figure}[H]
\centering
\centerline{\includegraphics[width=0.65\textwidth]{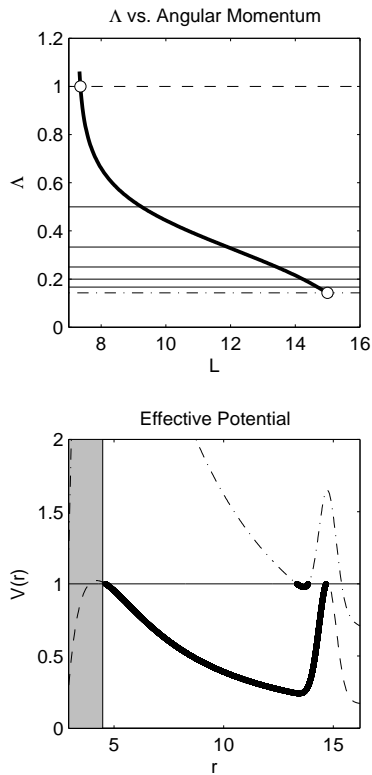}}
\caption{\textbf{Top} panel: $\Lambda$ as a~function of angular momentum $L$ is plotted for $E=\omega_\infty=1$ as the thick black line. On this panel we plot the value of $\Lambda=1$ as a~dashed line, $1/2$, $1/3$, $1/4$, $1/5$, $1/6$ as thin solid lines and $\Lambda=1/7$ as a~dash-dotted line. The intersections of the $\Lambda=1$ and $\Lambda = 1/7$ lines with $\Lambda(L)$ are marked as white discs. The effective potential corresponding to the angular momentum specified by these discs are plotted in the \textbf{lower} panel. The shaded region represents the interior of the CO and the heavy black lines are the effective potentials experienced by the ray. The $\Lambda=1$ and $\Lambda=1/7$ cases are represented by dashed and dash-dotted lines respectively.}
\label{figDelPhi}
\end{figure}

\begin{figure}[H]
\centering
\centerline{\includegraphics[width=0.85\textwidth]{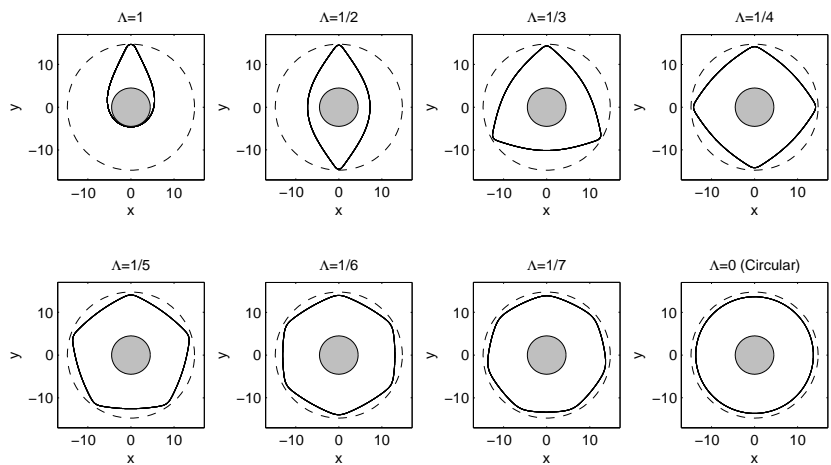}}
\caption{Periodic orbits with angular momentum $L$ found from $\Lambda=1$, $1/2$, $1/3$, $1/4$, $1/5$, $1/6$, $1/7$ and the stable circular orbit. The CO is the shaded disc, and the unstable circular orbit is marked with a~dashed line.}
\label{figPeriodics}
\end{figure}

  Bound ray trajectories admit periodic solutions with the appropriate angular momentum $L$ for a~given energy $E$. Consider the time elapsed between the closest distance of the orbit from the central CO, the periastron radius $r_\text{p}$ and the furthest distance, the apastron radius $r_\text{a}$. Let us define the angle accumulated as the integral of Equation \eqref{dphidr} for a~bound orbit,
\begin{equation}
\Delta \phi_\text{r} = 2 \int^{t(r_\text{a})}_{t(r_\text{p})} \frac{\text{d}\phi}{\text{dt}} \text{dt} = 2 \int^{r_\text{a}}_{r_\text{p}} \frac{\text{d}\phi}{\text{dr}} \text{dr}.
\end{equation}
Using Equation \eqref{dphidr} the right hand side gives
\begin{equation}
\Delta \phi_\text{r} = 2 \int^{r_\text{a}}_{r_\text{p}} \frac{\text{dr}}{r^2\left[ \frac{E^2}{L^2}n(r)^2 - \frac{A(r)}{r^2} \right]^\frac{1}{2}} .
\label{DelPhiDef}
\end{equation}
The quantity $\Delta \phi_\text{r}$ is the ratio of the radial and angular frequencies found from the Hamiltonian using action-angle variables \citep{levin},
\begin{equation}
\frac{\Delta \phi_\text{r}}{2\pi} = \frac{\omega_\phi}{\omega_\text{r}} = \Lambda.
\end{equation}
Periodic orbits have a~rational value of $\Lambda$. In the upper panel of Figure~\ref{figDelPhi} we calculate $\Lambda$ as a~function of $L$, holding $E=\omega_\infty=1$, via Equation \eqref{DelPhiDef}. We then numerically find the corresponding values of $L$ for which $\Lambda$ equals $1$, $1/2$, $1/3$, $1/4$, $1/5$, $1/6$ and $1/7$. The stable circular orbit has $\Lambda=0$. For $\Lambda=1$ (dashed line) and $\Lambda=1/7$ (dash-dotted line) we plot the corresponding effective potentials in the lower panel, and show the potential experienced by the bound orbit as a~thick black line. The~ratio $\Lambda$ is bounded in $L$ since a~bound orbit requires a~potential minimum. When $L$ is too low, no inner (stable) potential boundary exists between $R$ and $r_\text{ECS}$. For $L$ too high, orbits are not bound by the contribution to the effective potential from the shell. In Figure~\ref{figPeriodics} we plot the periodic orbits corresponding to these $\Lambda$ values.

\begin{figure}[H]
\centering
\centerline{\includegraphics[width=0.8\textwidth]{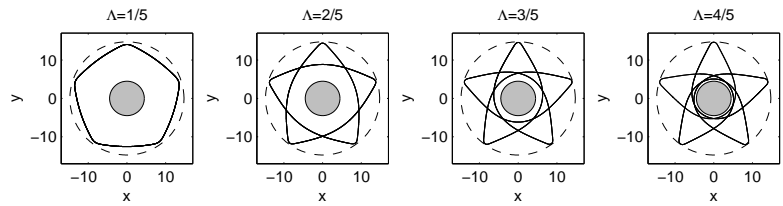}}
\caption{A family of periodic orbits with $L$ found from $\Lambda=1/5$, $2/5$, $3/5$ and $4/5$. The orbits reflect off of the plasma shell $5$ times, and the orbital path intersects itself $0$, $5$, $10$ and $15$ times, respectively. The CO is the shaded disc, and the unstable circular orbit is marked with a~dashed line.}
\label{figFamily}
\end{figure}

  For the periodic orbits the denominator of the rational $\Lambda$ values denote the number of deflections from the effective potential barrier provided by the plasma in a~complete orbit. The numerator describes the complexity of the orbit. For example, let $\Lambda=\alpha / \beta$ be a~rational number. The resulting orbit has $\beta$ reflections off of the plasma shell and the path intersects itself $(\alpha-1)\beta$ times. A~family of periodic orbits with $\Lambda=1/5$, $2/5$, $3/5$ and $4/5$ is plotted in Figure~\ref{figFamily} to demonstrate. These orbits each contain $5$ reflections from the plasma shell and feature $0$, $5$, $10$ and $15$ intersections of the orbital path.

\begin{figure}[H]
\centering
\centerline{\includegraphics[width=0.4\textwidth]{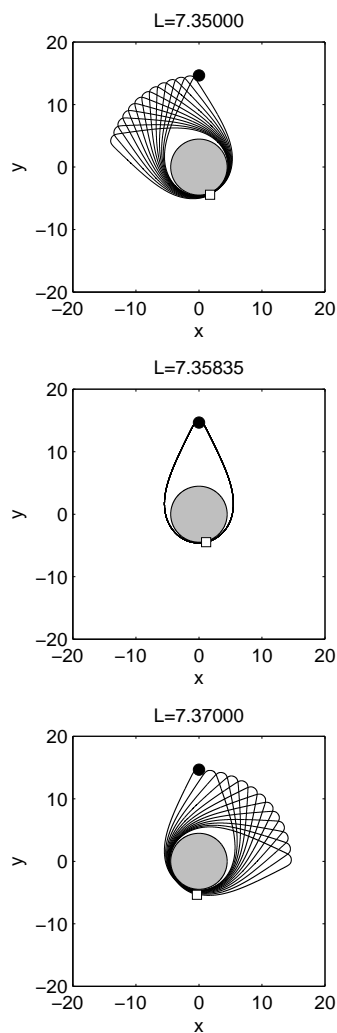}}
\caption{An example of precession as a~function of angular momentum for $\Lambda=1$ (\textbf{middle} panel). For $\Lambda>1$ we find precession in the counter-clockwise direction (\textbf{top} panel) and for $\Lambda<1$ the orbit precesses in the opposite sense (\textbf{bottom} panel). Initial positions are plotted as black dots and final positions are squares.}
\label{figPrecession}
\end{figure}

 In terms of angular momentum, the periodic configurations are preceded and followed by orbits with similar morphology but which precess in the opposite sense. For example, the $\Lambda=1$ periodic orbit is shown in the center panel of Figure~\ref{figPrecession}. We integrate the equations of motion until the coordinate time component reaches an~arbitrary value sufficient to provide several orbits of the CO. The initial position is marked as a~black dot and the final position is a~square. Given a~periodic orbit with $\Lambda=q$, a~lower angular momentum orbit with $\Lambda>q$ precesses in a~counter-clockwise manner (top panel), and an~orbit with $\Lambda<q$ precesses in the opposite sense (bottom panel).

\unskip

\unskip

\subsection{Frequency Windows}

  We define two particularly significant frequency ranges for rays that interact with the plasma atmosphere. The escape window (EW) is defined by the frequency range $\omega_{\infty 0}<\omega_\infty \leq \omega_{\infty+}$. Rays~in this frequency window can escape from the CO surface to reach an~observer at infinity, but the trajectories are strongly influenced by the presence of the plasma atmosphere. Below the EW a~second frequency band exists, $\omega_{\infty-} < \omega_{\infty} \leq \omega_{\infty 0}$, which we refer to as the anomalous propagation window (APW). In this frequency range, rays are trapped by the plasma. Rays in the APW that leave the surface of the CO travel to a~maximum altitude and then turn back to the CO surface. This trapping is analogous to the anomalous propagation of low-frequency radio waves in the atmosphere of the Earth. In addition to a~family of trapped rays emitted from the CO surface, the APW also includes a~set of rays that approach the the plasma atmosphere from the outside and are scattered off of it. The EW and APW were explored in \cite{rogersOrbits} for a~CO surrounded by a~cloud of plasma with a~power-law density.

\textcolor{black}{We define the asymptotic plasma frequency $\omega_{\infty \text{p}}$ in analogy with the ray frequency at infinity (Equation \eqref{pT}) using the redshift definition (Equation \eqref{redshiftDef}),}
\begin{equation}
\omega_{\infty \text{p}}=A(r)^\frac{1}{2}\omega_\text{p}(r),
\label{plasmaFreqInf}
\end{equation}
such that the propagation of radiation requires $\omega_\infty$ to exceed $\omega_{\infty \text{p}}$ along the path of a~ray. In fact, the quantity on the right hand side is simply the square root of the effective potential with $L=0$, for~which $r_\text{u} = r_\text{ECS}$.

\textcolor{black}{The limiting frequency for a~radially directed ray to escape is given by Equation \eqref{plasmaFreqInf} at $r_\text{ECS}$,}
\begin{equation}
\omega_{\infty 0} = \sqrt{V(r_\text{u})} = \lambda \sqrt{k N\left( r_\text{ECS} \right)} = \lambda \sqrt{k}.
\label{freqInf0}
\end{equation}
The condition on radially directed rays defines the floor of the EW and the APW ceiling. We will focus on each of the frequency ranges separately in the following sections.

\subsubsection{Ray Trapping}
\label{sec:trapping}

  Outgoing radially-directed rays with $\omega_{\infty} \leq \omega_{\infty 0}$ \textcolor{black}{cannot propagate through the plasma atmosphere.} Since the effective potential contains a~contribution from the angular momentum, turning~points exist for all rays with finite $L$,
\begin{equation}
\omega_\infty = \sqrt{V(r)} > \omega_{\infty \text{p}}.
\end{equation}
Therefore, provided $L \neq 0$, all rays with $\omega_{\infty} \leq \omega_{\infty0}$ are scattered. Rays with finite angular momentum that are emitted from the CO surface reach a~maximum height and return to the CO surface. \textcolor{black}{This trapping requires a~potential maximum outside the stellar surface, $V(R) \leq V(r_\text{u})$, and~thus we define the lower APW limit as}
\begin{equation}
\omega_{\infty -} = \sqrt{V(R)}.
\label{freqInfMinus}
\end{equation}
For incoming rays that approach the CO atmospheric shell externally, we write the angular momentum as $L=\omega_{\infty} b$ where $b$ is the impact parameter. For non-vanishing $b$, the approaching rays reach a~minimum distance before scattering from the plama atmosphere due to the presence of a~turning point and return to infinity.

\begin{figure}[H]
\centering
\centerline{\includegraphics[width=0.55\textwidth]{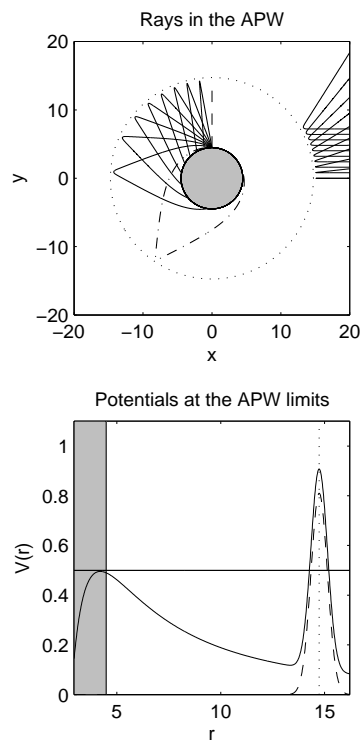}}
\caption{Trapped and scattered rays in the APW. The \textbf{upper} panel plots the paths of $10$ rays emitted from the CO surface, $R$, at $\phi=\pi/2$. These rays are evenly spaced between $L=0$ (dashed), and~$L_\text{max}$ (dashed-dotted). Also plotted are the paths of $10$ rays that approach the ECS externally from the $+x$ direction, and are reflected by the potential boundary. These external rays are plotted as the solid black curves. The ECS radius is the dotted circle. The \textbf{lower} panel shows the corresponding effective potential for the bound rays $L=0$ (dashed) and $0.99 L_\text{max}$ (solid). The square of the asymptotic frequency $\omega_\infty^2$ is the horizontal line. The shaded region is the CO interior, and the vertical dotted line is the ECS radius. \textcolor{black}{Both interior and exterior radially directed rays ($L=0$) have frequencies $\omega_\infty=\omega_{\infty p}$. These~rays, and~all rays with lower frequency, cannot propagate through the plasma.}}
\label{figAPW}
\end{figure}

We plot examples of rays in the APW in Figure~\ref{figAPW} and use $\omega_{\infty}=\sqrt{1/2}$, well below the APW ceiling $\omega_{\infty 0}=\lambda k^{1/2}=0.9$ for the choice of example parameters in our units. We plot the trajectories of ten rays launched from $R$ at position $\phi=\pi/2$. We vary the angular momenta of the rays from $0$ (dashed ray) to $0.99 L_\text{max}$ (dashed-dotted ray), where the maximum angular momentum is found by setting by relating the square of the frequency and the turning point at the surface,
\begin{equation}
L_\text{max} = \omega_\infty \frac{R}{A(R)^\frac{1}{2}}.
\end{equation}
At and above $L_\text{max}$, the ray is in a~bound orbit and only skims the surface tangentially at $R$. The~ECS radius is plotted as a~dotted circle. External rays that approach from the $+x$ direction are scattered by the plasma atmosphere. For these rays we used the same frequency and an~identical set of angular momenta, giving the impact parameters
\begin{equation}
b=\frac{L}{\omega_\infty}.
\end{equation}
The fiducial ray does not propagate in the plasma since $\omega_\infty=\omega_{\infty p}$ for the $L=0$ case. In the bottom panel we plot the effective potential for the $L=0$ ray (dashed curve) and the potential for a~ray with $0.99 L_\text{max}$ (solid curve). The horizontal line is $\omega_{\infty}^2$. The interior of the CO is the shaded region and the ECS radius is plotted as the vertical dotted line.

\subsubsection{Ray Escape}
\label{sec:escape}

  The angular momentum corresponding to the upper limit of the EW is generally found when
\begin{equation}
V(R)=V(r_\text{u}).
\label{LEW}
\end{equation}
Using the above relationship to find the angular momentum results in
\begin{equation}
L_{+}=k^\frac{1}{2}\left[ \frac{A(R)}{R^2} - \frac{A(r_\text{u})}{r_\text{u}^2}\right]^{-\frac{1}{2}}\left[A(r_\text{u})N(r_\text{u}) - A(R)N(R) \right]^{\frac{1}{2}}
\label{LEWGeneral}
\end{equation}
This expression simplifies for the case of a~gaussian shell, in which case we realistically assume the extent of the atmosphere is thin and does not significantly contribute at $r=R$, and that the ECS is sufficiently far from the stellar surface such that $r_\text{u} \approx r_\text{ECS}$. With these assumptions Equation~\eqref{LEWGeneral}~becomes
\begin{equation}
L_{+}=\lambda R \left[ \frac{k}{A(R) - \left[\frac{R \lambda (1-\lambda^2)}{2M} \right]^2} \right]^\frac{1}{2}
\end{equation}
which gives the asymptotic frequency
\begin{equation}
\omega_{\infty+}=\sqrt{V(R)} \approx A(R)^\frac{1}{2} \frac{L_{+}}{R}.
\label{EWTop}
\end{equation}
Rays with frequencies in the range $\omega_{\infty 0} < \omega_\infty < \omega_{\infty +}$ can escape the surface of the CO and are detectable by distant observers.

In the vacuum case the maximum impact parameter is given by a~ray escaping from an~emission point at $R$ with a~launch angle $\delta=\pi/2$ with respect to the radial direction to arrive at a~distant observer. The maximum impact parameter is given by
\begin{equation}
b_\text{max}=R \frac{n(R)}{A(R)^\frac{1}{2}}.
\label{impactParVac}
\end{equation}
The maximum impact parameter for a~ray in the EW with frequency $\omega_\infty$ differs from the vacuum case due to the presence of the unstable circular orbit, and is found using Equation \eqref{impactParVac} with $R$ replaced by $r_\text{u}$,
\begin{equation}
b_\omega=r_\text{u} \frac{n(r_\text{u})}{A(r_\text{u})^\frac{1}{2}}.
\end{equation}
The ray is launched at an~angle given by the ratio with the maximum impact parameter,
\begin{equation}
\delta_\omega = \sin^{-1} \left( \frac{b_\omega}{b_\text{max}} \right).
\end{equation}
The maximum impact parameter for a~ray in the EW is the solution for which the denominator of Equation \eqref{dphidr} vanishes. Expanding the denominator gives
\begin{equation}
r^3-b^2r+2Mb^2- r^3\frac{\omega_\text{p}^2}{\omega_\infty^2}+ 2Mr^2\frac{\omega_\text{p}^2}{\omega_\infty^2}=0.
\end{equation}
The minimum of this cubic function occurs at $r_\text{u}$, and the choice $b_\omega$ ensures a~single unique solution \cite{rogersOrbits}.

\begin{figure}[H]
\centering
\centerline{\includegraphics[width=0.65\textwidth]{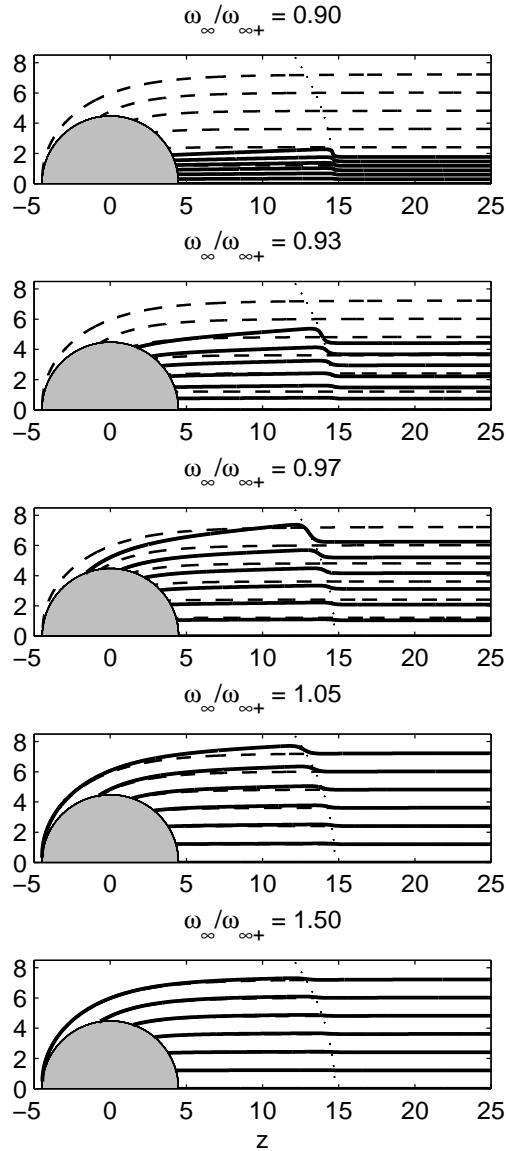}}
\caption{Trajectories of rays in the EW launched from the surface of the CO with frequency ratios $\omega_\infty/\omega_{\infty+}=0.90$ (\textbf{top} panel); $0.93$ (\textbf{second} panel) and $0.97$ (\textbf{third} panel); For rays higher than the EW, we also show the case for $\omega_\infty/\omega_{\infty+}=1.05$ (\textbf{fourth} panel) and $1.50$ (\textbf{bottom} panel).}
\label{figEW}
\end{figure}

  Beyond the EW, $\omega_\infty>\omega_{\infty+}$, rays experience trajectory modification between the surface of the CO and the atmospheric plasma shell but approximate the vacuum trajectories at greater distances. As the frequency is increased, the vacuum trajectories are recovered. We show the ray trajectories for a~variety of rays within and above the EW in Figure~\ref{figEW}. For our example parameters, we launch rays in the EW from the surface of the CO with the frequency ratio $\omega_{\infty}/\omega_{\infty+}$ for values of $0.90$, $0.93$, $0.97$. We also plot the cases for rays above the EW for frequency ratios $1.05$ and $1.50$. The highest frequency ratio appears very similar to the vacuum case.

\section{Discussion}
\label{sec:discussion}

  Let us estimate the minimum observable frequency $\omega_{\infty0}$ visible to a~distant observer. Since this quantity depends on the ratio of stellar to Eddington luminosity, we will estimate an~upper limit given by the plasma frequency (Equation \eqref{plasmaFreq}) at the ECS radius. We use the maximum mass density $\rho_0$ in the range $10^{-3}$ to $10^{-5}$ g$\cdot$cm$^{3}$ \cite{atmThick} to estimate the plasma frequency cut-off. Converting from plasma particle mass density to number density gives $N_{0}=\rho_{0}/(\mu m_p)$ where $m_p$ is the proton mass and $\mu=1/2$ the mean molecular weight. With these values we find $\omega_\text{p}$ between $\sim$6 and $\sim$60 THz, in~the mid to near infrared portion of the electromagnetic spectrum. However, since $\omega_{\infty-}$ depends on the value of $\lambda$, this estimate is an~upper limit. Moreover, for the choice of $N(r)$ used here, we~emphasize the effect of changing the plasma frequency constant $k$ amounts to a~simple re-scaling of the dynamical quantities that enter the equations of motion as discussed in Section \ref{sec:numerical}.

  Other effects may additionally reduce this estimate, for example if the atmospheric plasma particles have a~significant velocity spread the Lorentz factor modifies the effective plasma frequency~\cite{plasmaFreqLorentz}. Thus, we expect the observed frequency range to be sensitive to the effects of temperature in the atmospheric shell \cite{atmThin} and strong magnetic fields from the CO (neglected in previous studies). Including these conditions requires a~more complex dispersion relation (for examples, see~\cite{brodBland}).

  Our results show that while the high frequency radiation receives little modification from the vacuum case, the lower frequency components are dramatically modified, particularly for frequencies in the EW. The behaviour of these rays under the influence of both gravitation and the optical effects introduced by the plasma atmosphere significantly affect the appearance of the CO that a~distant observer would measure. This frequency-dependent view alters the observed pulse profile, an~effect which has been studied for COs surrounded by a~power-law plasma density \cite{rogers15, rogersProc, rogersOrbits}. The change in behaviour of the pulse profiles at frequencies in the EW may give an~analogous signature of the presence of the optically thin levitating atmospheric shell. With a~relatively high plasma frequency such effects may be feasible to observe.

  \textcolor{black}{We have considered non-rotating COs and have assumed spherical symmetry}. In practice, rotation significantly impacts pulse profiles, particularly through the Doppler effect \cite{cadeau07}. Including the effects of rotation would provide additional modifications to the basic spherically symmetric calculations detailed here.

\begin{figure}[H]
\centering
\centerline{\includegraphics[width=0.56\textwidth]{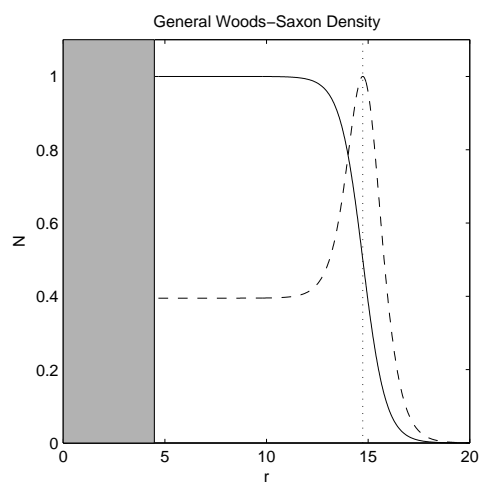}}
\caption{The general Woods-Saxon function. The dashed curve ($a=1$, $C=8$) shows a~density $N(r)$ with a~discrete maximum at $r_\text{ECS}$. This function does not vanish at the CO surface at $R$, shown by the shaded portion on the left of the figure. The solid curve ($a=1$, $C=0$) represents the extreme case in which the density does not monotonically decrease toward the star. Both of these functions produce regions of constant density above the stellar surface. The vertical dotted line denotes the position of~$r_\text{ECS}$.}
\label{fig:WoodsSaxon}
\end{figure}

  Finally, we demonstrate the generality of our results by considering an~extreme case. The~gaussian shell density has a~well-defined local maximum at $r_\text{ECS}$, so it is interesting to consider the resulting trajectories when this assumption does not hold and the shell is thick. We have constructed a~density profile based on the form of the generalized Woods-Saxon potential used in modeling nuclear interactions \cite{woodsSaxon}. The resulting density profile differs significantly from the gaussian shell scenario since the Woods-Saxon type density does not vanish at the stellar surface. Though these cases are not necessarily physically realistic, they provide an~interesting test of our results. The Woods-Saxon density is
\begin{equation}
 N(r)=\frac{a}{1+\exp\left(\frac{r-r_0}{\sigma} \right)} + \frac{C \exp\left( \frac{r-r_0}{\sigma} \right) }{\left[1+\exp\left( \frac{r-r_0}{\sigma} \right) \right]^2}
\end{equation}
where $a$ and $C$ are constants, and normalized to give a~maximum at $1$. We illustrate the potential for two example parameter sets. We plot $a=1$, $C=8$, and $r_0$ adjusted to produce a~maximum at $r_\text{u}=r_\text{ECS}$ when $C>1$,
\begin{equation}
r_0=r_\text{ECS} - \sigma \ln\left[ \frac{C-1}{C+1} \right]
\end{equation}
such that
\begin{equation}
 \left. \frac{\text{d}N(r)}{\text{dr}}\right|_{r_\text{ECS}} = 0.
\end{equation}
We also include the normalization condition $N_0=(C+1)^2/(4C)$. This potential gives a~non-vanishing contribution at $R$, in contrast with the gaussian shell example. In~addition, consider~the function with $a=1$, $C=0$ which does not have a~discrete maximum and instead produces a~shelf-like region of constant density above the stellar surface. The densities corresponding to these parameter sets are shown in Figure~\ref{fig:WoodsSaxon}.

Using the Woods-Saxon density, we find families of periodic and trapped orbits for each configuration. The $C=8$ density gives a~range of periodic orbits with $\sim$0.3 < $\Lambda$ < $\sim$0.8 and $\sim$7~<~$L$~<~$\sim$11.5 for $\omega_\infty=1$. The morphology of the periodic orbits in this set are qualitatively similar to those found for the gaussian shell solution, and differ only in the location of the turning points. When $C=0$ the Woods-Saxon density gives a~limited range of periodic orbits that are nearly circular with $\Lambda=4$ and $\Lambda=5$. Despite these differences, the qualitative features of the analysis given in terms of the orbital precession discussed in Section \ref{subsec:periodic} holds. Since the density is multiplied by $A(r)$ in the potential, both of these density functions produce a~maximum in the effective potential $r_\text{u}>R$ and therefore also have a~family of bound orbits. These examples show that our analysis is robust despite significant changes in the density profile and does not depend sensitively on the precise details of the assumed density profile of the plasma shell. We have also tested our conclusions for a~variety of density profiles that were based on the wave functions of the Hydrogen atom. These density profiles have a~local maximum, give a~finite contribution at the stellar surface, and are asymmetric about $r_\text{u}$ but show monotonic decrease as $r$ changes from the potential maximum. In all of these cases we also found periodic orbits that were analagous in morphology to those found using the gaussian shell. These potentials also produced trapped orbits that are analogous to those found in the gaussian case.

\section{Conclusions}
\label{sec:conclusions}

  Levitating shells of plasma above the surface of COs have been studied in both optically thin and thick cases \cite{atmThin, atmThick}. We have studied optically thin atmospheric shells around spherically symmetric COs with a~density maximum $N(r_0)$ well-separated from the stellar surface, $r_{0}>R$. Physical solutions require that $N(r)$ shows monotonic decrease in both the $r<r_0$ and $r>r_0$ directions. We find a~stable circular orbit at the potential minimum ($r_\text{s}$) and an~unstable circular orbit at the potential maximum ($r_\text{s} < r_\text{u}$). Existence of the stable circular orbit does not require any other conditions on the shell density, which can be asymmetric in $r$. We stress that the particular choice of the density distribution $N(r)$ does not dramatically affect these general results which hold for any shell-like density distribution with a~local maximum $r_0>R$.

  A~family of bound orbits exist for the effective potential between the unstable circular orbit $r_\text{u}$ and the stellar surface at $R$. These orbits assume a~variety of unique morphologies which we have categorized based on the ratio of radial and angular frequencies that arise from the Hamiltonian in action-angle coordinates, which we label $\Lambda$. For rational values of $\Lambda=\alpha/\beta$, we find non-precessing orbits. The denominator $\beta$ determines the number of turning points along the orbital path, and~the numerator describes the number of times the orbital path intersects itself ($\left[\alpha-1\right]\beta$). Orbits with values of $\Lambda$ slightly above and below a~given rational value produce orbits with similar morphologies, but~which precess in opposite directions.

  Two significant frequency windows exist in which low-frequency rays are strongly affected by the plasma shell. The APW is defined by the frequency range $\omega_{-} < \omega \leq \omega_{0}$, in which rays emitted from the stellar surface $R$ will be deflected by the potential boundary and return to the star. Rays~external to the potential boundary will likewise be reflected away from it. The EW is defined by the frequencies $\omega_{0} < \omega \leq \omega_{+}$, in which rays emitted from the CO surface are free to escape to distant observers. The~trajectories of these rays are strongly influenced by the plasma, which affects the observed appearance of the CO \cite{rogersOrbits}.

\vspace{6pt}

\acknowledgments{I acknowledge and thank Samar Safi-Harb for support through the Natural Sciences and Engineering Research Council of Canada (NSERC) Canada Research Chairs Program. I thank Andrew Senchuk for many stimulating conversations and suggestions for test density functions. \textcolor{black}{In addition, I also acknowledge both of the anonymous referees, who provided helpful and constructive input that refined the text.}}

\conflictofinterests{The author declares no conflicts of interest.}

\bibliographystyle{mdpi}

\renewcommand\bibname{References}

\end{document}